\documentclass[letterpaper, 12pt]{article}
\pdfoutput = 1

\usepackage{shortcuts}

\usepackage[margin = 2.5cm]{geometry}
\titleformat{\subsubsection}
  {\normalfont\normalsize \it}{\thesubsubsection}{1em}{}

\usepackage[nottoc]{tocbibind}

\begin{document}

\thispagestyle{empty}

\begin{flushright}
\texttt{BRX-TH-6694}
\end{flushright}

\begin{center}

\vspace*{0.08\textheight}

{\Large \textbf{Confinement and Flux Attachment}}

\vspace{1cm}

{\large \DJ or\dj e Radi\v cevi\'c}
\vspace{1em}

{\it Martin Fisher School of Physics\\ Brandeis University, Waltham, MA 02453, USA}\\ \medskip
\texttt{djordje@brandeis.edu}\\

\vspace{0.1\textheight}

\begin{abstract}
Flux-attached theories are a novel class of lattice gauge theories whose gauge constraints involve both electric and magnetic operators.  Like ordinary gauge theories, they possess confining phases.  Unlike ordinary gauge theories, their confinement does not imply a trivial gapped vacuum. 
This paper will offer three lessons about the confining phases of flux-attached $\Z_K$ theories in two spatial dimensions. First, on an arbitrary orientable lattice, flux attachment that satisfies a simple, explicitly derived criterion leads to a confining theory whose low-energy behavior is captured by an action of a general Chern-Simons form. Second, on a square lattice, this criterion can be solved, and all theories that satisfy it can be enumerated. The simplest such theory has an action given by a difference of two Chern-Simons terms, and it features a kind of subsystem symmetry that causes its topological entanglement entropy to behave pathologically. Third, the simplest flux-attached theory on a square lattice that does \emph{not} satisfy the above criterion is exactly solvable when the gauge group is $\Z_2$. On a torus, its confined phase possesses a twofold topological degeneracy that stems from a sum over spin structures in a dual fermionic theory. This makes this flux-attached $\Z_2$ theory an appealing candidate for a microscopic description of a U(1)$_2$ Chern-Simons theory.
\end{abstract}

\end{center}

\newpage

\section{Introduction and summary}

Chern-Simons (CS) theory is ubiquitous. It appears in connection to a bewildering array of phenomena in quantum field theory (QFT), see e.g.\ \cite{Deser:1982vy, Redlich:1983kn, Redlich:1983dv, Niemi:1983rq, Witten:1988hf, Polyakov:1988md, Witten:1988hc, Zhang:1988wy, Witten:1992fb, Zee:1995, Gopakumar:1998ki, Witten:2003ya, Aharony:2008ug, Giombi:2011kc, Closset:2012vp, Banerjee:2013mca, Hsin:2016blu}. This is why it is vexing that we possess no definition of CS theory that is precise, fully regularized, and manifestly quantum. Specifically, the issue is that there is no known quantum theory with a $D$-dimensional Hilbert space, $D < \infty$, from which one can obtain the salient features of CS via an explicit calculation, either exactly or at leading order in $1/D$.\footnote{There are many formulations of CS theory that give up one of the three desiderata listed here. For example, there exist various ways to rigorously define the algebra of operators and their expectation values in a CS theory \cite{Schwarz:1978cn, Atiyah:1988, Reshetikhin:1991tc,  Walker:1992, Freed:1992vw}, but none explicitly constructs a physical microscopic Hilbert space and a local Hamiltonian acting on it. On the physics side there exist rather detailed proposals for placing Abelian CS theories on discretized manifolds \cite{Frohlich:1988qh, Eliezer:1991qh, Alekseev:1994pa, Sun:2015hla}, but these approaches leave the target space unregularized. Finally, many fully regularized condensed matter systems are believed to give rise to CS theories (``chiral spin liquids'') at low energies, see e.g.\ \cite{Wen:1989yjx, Schroeter:2007, Yao:2007, Szasz:2020}, but we have no precise analytic description of this emergence.}

It is not unreasonable to ask for such a precise definition of a QFT. Many familiar theories can be defined this way. At least heuristically, this has been known since the early work on universality and the renormalization group, when low-energy regimes of various lattice models were shown to be well described by continuum QFTs \cite{Patashinsky:1966, Kadanoff:1966wm, Kadanoff:1969zz, Polyakov:1970xd, Kadanoff:1970kz, Wilson:1971bg, Wilson:1975id, Cardy:1996xt, Sachdev:2011}. Recent years have seen a renewed (if unsynchronized) effort to better understand this ``lattice-continuum correspondence'' and its limitations \cite{Mong:2014ova, OBrien:2017wmx, Milsted:2016jms, Milsted:2017csn, Zou:2019dnc, Zou:2019iwr, Radicevic:2019jfe, Radicevic:2019mle, Gorantla:2020xap, Gorantla:2021bda, Radicevic:1D, Radicevic:2D, Radicevic:3D}. Unfortunately, these new tools do not shed a completely illuminating light on the finite theory that underlies CS physics, except in $(0+1)$D \cite{Radicevic:3D}.

This paper will likewise not provide a definitive microscopic construction of any CS theory. Instead, it will highlight a large class of new Abelian lattice gauge theories that are natural candidates for such microscopic progenitors. These theories will be defined in the canonical formalism on two-dimensional spatial lattices, and their Hilbert spaces will be manifestly finite. They will also be sufficiently simple to allow explicit, controlled calculations.

The lattice gauge theories considered here will have nonstandard gauge constraints that feature \emph{flux attachment}. These are theories with local symmetries generated by products of Gauss operators and elementary Wilson loops. (Ordinary lattice gauge theories, on the other hand, have local symmetries generated just by Gauss operators.) Such theories play a key r$\hat{\trm o}$le in generalizing the Jordan-Wigner transformation to higher dimensions \cite{Chen:2017fvr, Chen:2018nog, Radicevic:2018okd}. In particular, in $(2 + 1)$D the simplest flux-attached gauge theory with gauge group $\Z_2$ is exactly dual to a spinless complex fermion.

Enshrining flux attachment in gauge constraints ensures that in the topological phase the Lagrangian density contains a CS-like term $A^0 F^{12}$ \cite{Chen:2017fvr}. However, this is where the similarity with CS theory seems to stop. This paper will show that further hallmarks of CS arise when the flux-attached gauge theory is in the \emph{confined} phase.

It may seem surprising that CS theory might emerge from a confining regime of a lattice gauge theory. After all, standard lore says that the vacuum in the confined phase is trivial, with exactly zero electric flux on each lattice link, so confinement should be distinct from the phase in which CS dominates \cite{Affleck:1989qf, Fradkin:1990xy}. Luckily, standard lore breaks down in the presence of nonstandard gauge constraints \cite{tHooft:1981bkw, Cardy:1981qy, Cardy:1981fd, Girvin:1987fp, Rey:1989jg, Diamantini:1994xr}. The presence of flux attachment means that electric operators on links are not gauge-invariant. Instead, gauge invariance is achieved by dressing electric operators with Wilson lines.\footnote{
 In an ordinary gauge theory in $(2+1)$D, an electric operator creates a pair of magnetic flux excitations on the two plaquettes touching its link. In a flux-attached theory, magnetic flux excitations are electrically charged, and so a gauge-invariant operator that creates a pair of magnetic fluxes must also create a line of electric flux between this pair of magnetic fluxes.
}
The confinement of such dressed operators relates electric fluxes to vector potentials instead of merely setting each electric flux to zero. This kind of relation is a feature of CS theory; this is why the present lattice approach is promising.

The remainder of this note consists of three parts. Each is dedicated to fleshing out one specific idea related to confinement in flux-attached theories.

\textbf{Section \ref{sec arbitrary lattice}} contains all the formal background. It introduces flux-attached theories on arbitrary lattices in a way that allows for more general flux attachment rules than originally considered in \cite{Chen:2017fvr}. It then precisely defines the confining phase for these theories. An important lesson is that all dressed electric operators can be simultaneously diagonalized only for some choices of flux attachment rules \cite{Radicevic:3D}. Theories with such flux attachment are called \emph{fully confinable}. Other theories are merely \emph{partially confinable}: only a subset of dressed electric operators can be confined at once. Fully confined flux-attached theories will be shown to have actions that look like the discretizations of U(1)$_K$ CS actions proposed in \cite{Sun:2015hla}.

\textbf{Section \ref{sec square}} specializes to the square lattice and explicitly finds all rules of flux attachment that let a theory be fully confinable. The simplest fully confinable $\Z_K$ theory is then studied in some detail. Its action is shown to be a difference of two CS actions; despite superficial similarities, this theory therefore does not give rise to CS in the confined phase. Nevertheless, this confined theory has some interesting properties reminiscent of fracton physics \cite{Nandkishore:2018sel}. In particular, this will prove to be a simple system whose topological entanglement entropy receives spurious contributions due to subsystem symmetries \cite{Williamson:2018zig}.

\textbf{Section \ref{sec part conf}} then shifts focus to partially confinable theories on the square lattice. In this case, even for the simplest choice of flux attachment rules, a Hamiltonian built out of dressed electric operators is not a sum of commuting projectors, and so it cannot be solved easily. However, for gauge group $\Z_2$ this theory can be solved by bosonization \cite{Chen:2017fvr}. This exact solution reveals a twofold topological degeneracy that stems from the sum over spin structures in the dual fermion picture. This suggests that the strongly coupled $\Z_2$ flux-attached theory, once supplemented by time reversal-breaking mass terms, is a promising candidate for a microscopic precursor of U(1)$_2$ CS theory.

\newpage

\section{Confining flux-attached theories on arbitrary lattices} \label{sec arbitrary lattice}

\bt{Definitions.} Consider an orientable $d = 2$ lattice $\Mbb$. A general one-chain $c$ is a linear combination of links of form
\bel{
  c \equiv \sum_{\ell \in \Mbb} \sigma^\ell_c\, \ell.
}
Importantly, to each site $v$ and plaquette $f$ one can associate the one-chains
\bel{\label{def bndry ops}
  \del_{-1} v \equiv \sum_{\ell \in \Mbb} \sigma^\ell_v \, \ell,
   \quad
  \del f \equiv \sum_{\ell \in \Mbb} \sigma^\ell_f \, \ell.
}
The coefficients $\sigma^\ell_{v/f} \in \{0, \pm 1\}$ will be chosen in the usual way, by first assigning a direction to each link. Then $\sigma_v^\ell = \pm 1$ differentiates whether $\ell$ ``flows'' out of or into $v$, and $\sigma_f^\ell = \pm 1$ tells whether $\ell$ ``wraps'' around $f$ in a counterclockwise or a clockwise direction. In each case, the coefficient is zero if $v \not\subset \ell$ or $\ell \not\subset f$.

A $\Z_K$ gauge theory on $\Mbb$ features shift and clock operators, $X_\ell$ and $Z_\ell$, defined on links. A product of operators associated to an arbitrary one-chain is defined as
\bel{\label{def prod over chains}
  \prod_{\ell \in c} \O_\ell
   \equiv
  \prod_{\ell \in \Mbb} \O_\ell^{\sigma^\ell_c}.
}
Two familiar operators associated to sites and plaquettes are defined as products over the special one-chains $\del_{-1}v$ and $\del f$. These are Gauss operators and elementary Wilson loops,
\bel{
  G_v \equiv \prod_{\ell \in \del_{-1}v} X_\ell
   \quad \trm{and} \quad
  W_f \equiv \prod_{\ell \in \del f} Z_\ell.
}

The familiar fact that Gauss operators and Wilson loops commute with each other can be more generally expressed as follows. The group commutator of operators $\O_c \equiv \prod_{\ell \in c} X_\ell$ and $\O'_{c'} \equiv \prod_{\ell' \in c'} Z_{\ell'}$ is
\bel{
  \O_c \O'_{c'} = \greek w^{\sum_{\ell \in \Mbb} \sigma^\ell_c \sigma^{\ell}_{c'}} \O'_{c'} \O_c,
}
where $\greek w \equiv \e^{\i \, \d A} = \e^{2\pi\i/K}$. In the particular case when $c = \del_{-1}v$ and $c' = \del f$, the commutator is controlled by the quantity
\bel{
  \sum_{\ell \in \Mbb} \sigma^\ell_v \sigma^\ell_f.
}
If $v\not\subset f$, every term in this sum is zero. If $v \subset f$, the only nonzero terms come from links that satisfy $v \subset \ell \subset f$. There are always exactly two such links, and the convention of assigning $\sigma$'s based on directed links ensures that the contributions from these links will cancel. Thus, for any $v$ and $f$, $G_v$ and $W_f$ commute.

%The remainder of this Section will use this formalism to derive general conditions under which flux-attached gauge theories can be described by Chern-Simons actions in the confining regime.

Flux-attached gauge theories have local symmetries generated by the generalized Gauss operators
\bel{
  \G_v \equiv G_v^p W_{a(v)}, \quad 1 \leq p < K,
}
where $W_{a(v)}$ is the product of elementary Wilson loops over the two-chain
\bel{\label{def a(v)}
  a(v) \equiv \sum_\alpha q_\alpha f_\alpha(v), \quad 1 \leq q_\alpha < K, \quad \trm{gcd}(p, q_\alpha) = 1.
}
This two-chain contains the key information about flux attachment. Imposing the constraint $\G_v = \1$ instead of the usual $G_v = \1$ results in a gauge theory in which a magnetic flux at any plaquette $f_\alpha(v)$ comes with an attached electric charge at site $v$.\footnote{Roughly speaking, a unit magnetic flux at $f_\alpha(v)$ gives rise to  $q_\alpha/p$ units of electric charge at $v$. If $q_\alpha/p$ is not an integer, a state with a single unit of flux at $f_\alpha(v)$ cannot be gauge-invariant. Instead, since $\trm{gcd}(p, q_\alpha) = 1$, magnetic fluxes in gauge-invariant states will come in multiples of $p$.}

It would actually be more precise to say that, in a flux-attached theory with constraints $\G_v = \1$, a magnetic flux at a plaquette $f$ comes with electric charges at all sites $v$ for which $f \in a(v)$. These sites will be referred to as \emph{anchors} of the plaquette $f$. A gauge-invariant shift operator $\X_\ell$ must move not only the magnetic flux across the link $\ell$, which is what $X_\ell$ does in an ordinary gauge theory; it must also move all the attached electric charges. The latter task is accomplished by Wilson lines that connect anchors of the two plaquettes separated by $\ell$. Thus the general form of this shift operator is
\bel{
  \X_\ell = X_\ell^p W_{c(\ell)},
}
where $c(\ell)$ is a one-chain that connects anchors of plaquettes that share the link $\ell$, and $W_{c(\ell)}$ is the product of clock operators $Z_\ell$ over this one-chain.

On a general lattice there is no canonical choice for $c(\ell)$. The only requirements it obeys come from demanding gauge invariance, i.e.\ that $\X_\ell$ commute with $\G_v$. This translates to the condition
\bel{\label{g inv condition}
  \greek w^{p \sum_{\ell' \in \Mbb} \sigma^{\ell'}_v \sigma^{\ell'}_{c(\ell)} - p \sigma^\ell_{a(v)}} = 1
}
for all $\ell$ and $v$. Here the coefficients $\sigma^\ell_{a(v)}$ come from the one-chain $\del a(v) = \sum_{\ell \in \Mbb} \sigma^{\ell}_{a(v)} \ell$ and can be expressed as $\sigma^\ell_{a(v)} = \sum_\alpha q_\alpha \sigma^\ell_{f_\alpha(v)}$ in terms of the data featured in \eqref{def a(v)}.

The simplest way to satisfy \eqref{g inv condition} is to demand
\bel{\label{g inv condition pert}
  \sigma^\ell_{a(v)} = \sum_{\ell' \in \Mbb} \sigma^{\ell'}_v \sigma^{\ell'}_{c(\ell)} \quad \trm{for all }v, \, \ell \in \Mbb.
}
This is a ``perturbative'' condition for gauge invariance. The more general (``nonperturbative'') condition would demand that the exponent in \eqref{g inv condition} be a nonzero multiple of $K$.

The two sides of \eqref{g inv condition pert} are nonzero only when $v$ is an anchor of exactly one of the two plaquettes bordering $\ell$. Then the sum on the r.h.s.\ picks out the link that hosts the Wilson line connecting $v$ to the anchor of the other plaquette bordering $\ell$. This condition thus relates the coefficient with which that particular Wilson line appears in $\X_\ell$ to the charges $q_\alpha$ that figure in the generalized Gauss operator $\G_v$.

\noindent \bt{Confinement.} A few flux-attached gauge theories have the remarkable property that all the operators $\X_\ell$ commute with each other \cite{Radicevic:3D}. Such theories will be called \emph{fully confinable} because they may possess a phase characterized by the constraint $\X_\ell = \1$ on every link. Other flux-attached theories will be called \emph{partially confinable} because only a subset of operators $\X_\ell$ can be simultaneously constrained. Both fully and partially confinable theories can have Hamiltonians that do not realize either kind of confinement at any value of the coupling.

The condition that any two operators $\X_{\ell}$ and $\X_{\ell'}$ commute can be expressed as
\bel{\label{conf condition}
  \greek w^{p \sigma^\ell_{c(\ell')} - p \sigma^{\ell'}_{c(\ell)} } = 1
  \quad \trm{for all }\ell, \, \ell' \in \Mbb.
}
As with gauge invariance, the simplest way to realize this is ``perturbatively,'' by demanding
\bel{\label{conf condition pert}
  \sigma^{\ell'}_{c(\ell)} = \sigma^\ell_{c(\ell')}\quad \trm{for all }\ell, \, \ell' \in \Mbb.
}
This condition is quite restrictive, as will become apparent in the next Section.

For now, assume that the rules for flux attachment are chosen such that the condition \eqref{conf condition} is satisfied. The simplest fully confining Hamiltonian is
\bel{\label{def H}
  H = \sum_{\ell \in \Mbb} \left[ 2 - \X_\ell - \X_\ell\+ \right].
}
One can view $H$ as the $g \rar \infty$ limit of the flux-attached Kogut-Susskind Hamiltonian \cite{Kogut:1974ag}
\bel{\label{def H KS}
  H\_{KS} =
  \frac{g^2}{2(\d A)^2} \sum_{\ell \in \Mbb}
    \left[ 2 - \X_\ell - \X_\ell\+ \right]
   +
  \frac1{2g^2} \sum_{f \in \Mbb}
    \left[ 2 - W_f - W_f\+ \right].
}
The ground states of \eqref{def H} satisfy $\X_\ell = \1$ on every link. (A basic fact about this theory's self-consistency is that the product $\prod_{\ell \in \del_{-1}v} \X_\ell$ must equal the Gauss operator $\G_v$ up to a complex phase, and so the Gauss law constraint need not be imposed separately.)

In an ordinary gauge theory, imposing the confinement constraints $X_\ell^p = \1$ results in a Hilbert space of dimension $p^{|\Mbb|_1}$, where $|\Mbb|_1$ is the number of links in $\Mbb$.  The situation is very different in a flux-attached theory, even at full confinement. When $p \neq 1$, the dimensionality of the fully confined Hilbert space sensitively depends on the rules of flux attachment and the geometry of the lattice. This will be explicitly demonstrated in the next Section.

\noindent \bt{Actions.} This parametric sensitivity of the fully confined subspace is, perhaps surprisingly, captured by a relatively simple Euclidean action. When $K \gg 1$, this action belongs to a class of discrete-spacetime actions proposed by \cite{Sun:2015hla} as lattice regularizations of CS theory.

Consider the partition function associated to the low-temperature behavior of the theory \eqref{def H},
\bel{
  \Zf = \Tr \, \e^{-\beta H}, \quad \beta \gg 1.
}
The action can be found using the usual transfer matrix procedure, i.e.\ by writing
\bel{
  \e^{-\beta H} = \prod_{\tau \in \Sbb} \e^{-\d\tau H},
   \quad
  \d\tau \equiv \frac\beta{N_0} \ll \beta,
   \quad
  \Sbb \equiv \{\d\tau, 2\d\tau,\ldots, \beta\},
}
and inserting a decomposition of unity $\1 = \sum_{A_\tau} \qproj{A_\tau}{A_\tau}$ between each pair of $\e^{-\d\tau H}$'s. This gives
\bel{\label{part fn}
  \Zf
   =
  \sum_{\{A_\tau\}}
  \prod_{\tau \in \Sbb}
    \qmat{A_{\tau + \d\tau}} {\e^{-\d\tau H}} {A_\tau}.
}
Here $\qvec {A_\tau} \equiv \prod_{\ell \in \Mbb} \qvec {A_{\ell, \tau}}$ is the eigenstate of all clock operators $Z_\ell$ with eigenvalues $\e^{\i A_{\ell, \tau}}$.

The large-$\beta$ limit can then be implemented by replacing each $\e^{-\d\tau H}$ by a projector $\trm P\_{conf}$ to the confined subspace. This projector can be written as\footnote{
 The last expression assumes that $X_\ell^p$ and $W_{c(\ell)}$ commute for all $\ell$, which will be true in all examples of interest. It is straightforward to modify the results to account for a violation of this assumption if a concrete application demands it.}
\bel{
  \trm P\_{conf}
   =
  \prod_{\ell \in \Mbb} \left(
    \frac1K \sum_{m_{\ell} = 1}^K \X_{\ell}^{m_\ell}
  \right)
   =
  \frac1{K^{|\Mbb|_1}}
  \sum_{\{m\}}
    \prod_{\ell \in \Mbb} X_\ell^{p m_\ell} W_{c(\ell)}^{m_\ell}.
}
Each matrix element in \eqref{part fn} then becomes
\bel{\label{mat elem}
  \qmat{A_{\tau + \d\tau}} {\trm P\_{conf}} {A_\tau}
   =
  \frac1{K^{|\Mbb|_1}}
  \sum_{\{m_\tau\}}
    \e^{\i \left(L_1[A_\tau, m_\tau] + L_2[m_\tau] \right)}
    \qprod{A_{\tau + \d\tau}}{A_\tau - p\, m_\tau \, \d A},
}
with
\bel{
  L_1[A, m]
   \equiv
  \sum_{\ell \in \Mbb}
    m_\ell A_{c(\ell)},
   \quad
  L_2[m]
   \equiv
  -\frac12 p\, \d A \sum_{\ell \in \Mbb}
    m_{\ell}\, m_{c(\ell)}.
}
(The sum over a one-chain is defined as $A_{c(\ell)} \equiv \sum_{\ell' \in c(\ell)} A_{\ell'} \equiv \sum_{\ell' \in \Mbb} \sigma_{c(\ell)}^{\ell'} A_{\ell'}$, in analogy to the product \eqref{def prod over chains}.) The Lagrangian $L_1$, linear in the Lagrange multipliers $m_\ell$, simply comes from applying $W^{m_\ell}_{c(\ell)}$ to the vector $\qvec{A_\tau}$. The Lagrangian $L_2$, quadratic in the $m_\ell$'s, comes from taking into account the shifts of $A_{\ell, \tau}$ due to the action of $X_\ell^{pm_\ell}$ when applying $W_{c(\ell)}^{m_\ell}$. To get this simple expression for $L_2$ it is necessary to use the condition \eqref{conf condition pert} for full confinement.

The scalar product in \eqref{mat elem} imposes the constraint
\bel{
  m_{\ell, \tau}
   =
  \frac1{p \, \d A} (A_{\ell, \tau} - A_{\ell, \tau + \d\tau})
   \equiv
  - \frac{\d\tau}{p \, \d A} (\del_0 A)_{\ell, \tau}.
}
Recalling that $\d A = 2\pi/K$, this means that the Lagrangians $L_1$ and $L_2$ ultimately yield actions
\gathl{\label{def S CS full conf}
  S_1[A]
   \equiv
  - \frac K{2\pi p}
  \sum_{\tau \in \Sbb} \sum_{\ell \in \Mbb}
    \d\tau\, A_{c(\ell), \tau} \, (\del_0 A)_{\ell, \tau} ,
   \\
  S_2[A]
   \equiv
  - \frac K{4\pi p}
  \sum_{\tau \in \Sbb} \sum_{\ell \in \Mbb}
    (\d\tau)^2\, (\del_0 A)_{c(\ell), \tau}\, (\del_0 A)_{\ell, \tau}.
}
These are the actions that figure in the partition function path integral,
\bel{
  \Zf
   =
  \frac1{K^{N_0 |\Mbb|_1}}
  \sum_{\{A\}}
    \e^{\i\left(S_1[A] + S_2[A]\right)}.
}

The one-derivative action $S_1$ bears a striking resemblance to the continuum action
\bel{
  \frac{\kappa}{4\pi} \int_{\Sbb} \d\tau \int_{\Mbb}
    \d^2 \b x\, \left(A^2\_c(\b x, \tau)\, \del_0 A^1\_c(\b x, \tau) - A^1\_c(\b x, \tau)\, \del_0 A^2\_c(\b x, \tau)\right)
}
that is obtained by fixing the gauge $A^0\_c(\b x, \tau) = 0$ in the usual CS action,
\bel{
  \frac{\kappa}{4\pi} \int_{\Sbb} \d\tau \int_{\Mbb} \d^2 \b x
    \, \epsilon^{\mu\nu\lambda} A^\mu\_c(\b x, \tau)\, \del_\nu A^\lambda\_c(\b x, \tau).
}
It would be premature to claim victory, however. Section \ref{sec square} will study a specific flux-attached gauge theory to show that this resemblance does \emph{not} perfectly hold up under scrutiny.

It is also possible to obtain a CS-like action which has not been gauge-fixed. Assuming that $\X_\ell = \1$ implies the Gauss law $\G_v = \1$, one may simply replace $\trm P\_{conf}$ by $\trm P\_{Gauss} \trm P\_{conf}$, where
\bel{
  \trm P\_{Gauss}
   =
  \prod_{v \in \Mbb} \left(
    \frac1K \sum_{n_v = 1}^K \G_v^{n_v}
  \right)
   =
  \frac1{K^{|\Mbb|_0}}
  \sum_{\{n\}}
    \prod_{v \in \Mbb} G_v^{p n_v} W_{a(v)}^{n_v}
}
projects to the $\G_v = \1$ sector. Summing over $n_v$ is guaranteed not to change $\Zf$. However, keeping $n_v$ around and setting $A_v^0 \,\d\tau \equiv n_v \, p \, \d A$ replaces the one-derivative action by
\bel{\label{def S1 CS full conf}
  S_1[A, A^0]
   \equiv
  - \frac K{2\pi p}
  \sum_{\tau \in \Sbb} \d\tau
  \Big[
    \sum_{\ell \in \Mbb}
      A_{c(\ell), \tau} \, \left((\del_0 A)_{\ell, \tau} - (\delta A^0)_{\ell, \tau} \right)
     +
    \sum_{v \in \Mbb}
      A_{v, \tau}^0 B_{a(v), \tau + \d\tau}
  \Big].
}
Here $\delta$ is the coboundary operator that takes a zero-form like $A^0_v$ to $(\delta A^0)_\ell \equiv - \sum_{v \in \Mbb} \sigma^\ell_v A^0_v$, and $B_{a(v)}$ is the sum of the two-form $B_f \equiv (\delta A)_f$ over the two-chain $a(v)$ from \eqref{def a(v)}.

A similar calculation can be carried out to obtain the action describing the low-tem\-per\-a\-ture thermodynamics of partially confined theories. Partial confinement amounts to imposing the constraints
\bel{
  \X_\ell = \1, \quad \ell \in C,
}
where $C$ is a subset of links such that the condition \eqref{conf condition} is fulfilled for any $\ell, \ell' \in C$. (There is generally no unique, or even canonical, choice of $C$.) Heuristically, for a generic choice of flux attachment rules, $C$ will contain at most half the links in $\Mbb$.  There are not enough constraints here to induce the Gauss law at every site. Inserting the projectors $\trm P\_{Gauss}$ is thus necessary when constructing the path integral for the partially confined sector.

The action that describes the partially confined regime associated to the subset $C$ can be derived using the same steps as above. The resulting one- and two-derivative actions are
\gathl{\label{def S CS part conf}
  S_1[A, A_0]
   =
  -\frac K{2\pi p}
  \sum_{\tau \in \Sbb} \d\tau \Big[
    \sum_{\ell \in C} A_{c(\ell), \tau} \left((\del_0 A)_{\ell, \tau} - (\delta A^0)_{\ell, \tau} \right)
     +
    \sum_{v \in \Mbb} A_{v, \tau}^0 B_{a(v), \tau + \d\tau}
  \Big],
   \\
  S_2[A, A^0]
   =
  -\frac K{4\pi p}
  \sum_{\tau \in \Sbb} (\d\tau)^2 \sum_{\ell, \ell' \in C}
    \sigma^{\ell}_{c(\ell')}
    \left((\del_0 A)_{\ell, \tau} - (\delta A^0)_{\ell, \tau}\right)
    \left((\del_0 A)_{\ell', \tau} - (\delta A^0)_{\ell', \tau}\right),
}
subject to
\bel{\label{constraint part conf}
  (\del_0 A)_{\ell, \tau} - (\delta A^0)_{\ell, \tau} = 0, \quad \ell \notin C.
}
This constraint comes from the inner product $\qprod{A_{\ell, \tau + \d\tau} - p (\delta n)_{\ell,\tau} \, \d A}{A_{\ell,\tau}}$ on all links that do \emph{not} belong to $C$. Its presence makes it possible to replace the sums over $\ell \in C$ by sums over $\ell \in \Mbb$ in the actions \eqref{def S CS part conf}. As a consequence, the actions of partially and fully confined theories are actually \emph{equal}. The only difference is that the functional integral in the partially confined theory has fields obeying the additional constraint \eqref{constraint part conf}. Intuitively, this constraint simply sets some of the Lagrange multipliers $m_{\ell, \tau}$ to zero, meaning that some of the confinement constraints $\X_\ell = \1$ are \emph{not} enforced.

Finally, it may be useful to explicitly compare the one-derivative actions in \eqref{def S1 CS full conf} and \eqref{def S CS part conf} to the actions that describe the \emph{topological}, $g \rar 0$ phase of a flux-attached gauge theory \eqref{def H KS}. This phase is described by the constraints $\G_v = \1$ and $W_f = \1$ on all sites $v$ and faces $f$. These constraints are equivalent to $G_v^p = \1$ and $W_f = \1$, so for $p = 1$ this is just the ordinary topological phase of a $\Z_K$ gauge theory. Its action can be written in the usual BF form
\bel{
  S\_{BF}[A, A^0] = -\frac{K}{2\pi p} \sum_{\tau \in \Sbb} \d\tau \sum_{f \in \Mbb} A^0_{f, \tau} (\delta A)_{f, \tau}
}
with the constraint $(\del_0 A)_{\ell,\tau} \d\tau = 0 \, \trm{mod}\, \frac{2\pi p}K$. This is analogous to the $A^0_{v, \tau} B_{a(v), \tau + \d\tau}$ term in the actions $S_1[A, A^0]$, even though the microscopic interpretation of $A^0$ is different. Confinement introduces a further coupling of $A^0$ and $A$ and brings one closer to CS.

\section{Full confinement on a square lattice} \label{sec square}

\begin{comment}

  lattice notation

  example actions and associated GSDs

  smoothing (separate section if I have something to say?)

\end{comment}

The conditions \eqref{g inv condition} and \eqref{conf condition} that lead to the CS-like action \eqref{def S1 CS full conf} are rather unwieldy on a general lattice $\Mbb$. This Section will first simplify matters by specializing to the square lattice. A specific fully confining flux-attached theory will then be studied in some detail.

\noindent \bt{Generalities.} Every site of an $N_1 \times N_2$ square lattice $\Mbb$ can be labeled by a pair of integers
\bel{
  \b x \equiv (x^1, x^2), \quad 1 \leq x^i \leq N_i.
}
The unit vectors are
\bel{
  \b e_1 \equiv (1, 0), \quad \b e_2 \equiv (0, 1).
}
For now, assume that the lattice is periodic, so that $\b x + N_i \b e_i \equiv \b x$ for $i \in \{1, 2\}$.

All links (i.e.\ unordered pairs of adjacent sites) can then be labeled by a site and a direction, so that
\bel{
  \{\b x, \b x + \b e_i\} \equiv (\b x, i).
}
The links will be oriented so that the coefficients $\sigma^\ell_v$ from \eqref{def bndry ops} are, for all $\b x$ and $i$,
\bel{
  \sigma^{(\b x, i)}_{\b x} = +1, \quad \sigma^{(\b x  - \b e_i, i)}_{\b x} = -1.
}
An operator $\O_\ell$ on link $\ell \equiv (\b x, i)$ will henceforth be denoted $\O_{\b x}^i$.

Finally, a plaquette with vertices $\{\b x, \b x + \b e_1, \b x + \b e_2, \b x + \b e_1 + \b e_2\}$ will simply be labeled by $\b x$. The coefficients $\sigma^\ell_f$ from \eqref{def bndry ops} will then be
\bel{
  \sigma^{(\b x, 1)}_{\b x} = \sigma^{(\b x + \b e_1, 2)}_{\b x} = +1,
   \quad
  \sigma^{(\b x + \b e_2, 1)}_{\b x} = \sigma^{(\b x, 2)}_{\b x} = -1.
}
The potential confusion about whether $\sigma^{(\b x, i)}_{\b y}$ refers to $\sigma^\ell_v$ or $\sigma^\ell_f$ will be avoided by substituting the values of these coefficients before writing any further formul\ae.

To specify the generalized Gauss operators, pick an integer $p$ and a set $\{(q_\alpha, \b r_\alpha)\}_\alpha$ so that
\bel{\label{def G general}
  \G_{\b x} = G_{\b x}^p \prod_\alpha W_{\b x + \b r_\alpha}^{q_\alpha}.
}
(The original flux-attached theory on the square lattice \cite{Chen:2017fvr} had $p = 1$, $q = 1$, and $\b r = \b 0$.) The ``perturbative'' condition \eqref{g inv condition pert} for gauge-invariance of operators $\X_{\b x}^i = (X^i_{\b x})^p W_{c(\b x, i)}$ is then
\bel{\label{g inv condition pert square}
  - \sum_{\alpha,\, j}
    q_\alpha \epsilon^{ij}
    \del_{y^j} \delta_{\b x,\, \b y + \b r_\alpha}
   =
  \sum_j \del_{y^j} \sigma^{(\b y - \b e_j, j)}_{c(\b x, i)},
}
where the discrete derivatives are defined so that $\del_{y^j} f(\b x, \b y) = f(\b x, \b y + \b e_j) - f(\b x, \b y)$.

%The relation can be informally recorded as $\nabla_{\b y} \sigma^{\b y}_{c(\b x, i)} = - \sum_\alpha q_\alpha \nabla_{\b y} \times \delta_{\b x - \b r_\alpha, \b y}$.
Any set of coefficients $\sigma^{(\b y, j)}_{c(\b x, i)}$ that satisfies the relation \eqref{g inv condition pert square} encodes a set of Wilson lines that connect anchors of the plaquettes on either side of the link $(\b x, i)$. As explained in Section \ref{sec arbitrary lattice}, these Wilson lines serve to transport the electric charge that must move together with the magnetic flux that is transported by the $X^i_{\b x}$ operators in $\X^i_{\b x}$. The canonical solution to this relation comes from ``cancelling the derivatives'' on both sides, which gives
\bel{\label{canonical flux attachment}
  \sigma_{c(\b x, i)}^{(\b y, j)} = - \sum_\alpha q_\alpha \epsilon^{ij} \delta_{\b x,\, \b y + \b r_\alpha + \b e_j}.
}
Other solutions can be obtained by adding ``divergenceless'' fields to $\sigma^{(\b y, j)}_{c(\b x, i)}$. This corresponds to adding loops to the connecting Wilson lines --- or, in other words, to choosing different paths that connect the same anchors on the two sides of $(\b x, i)$.

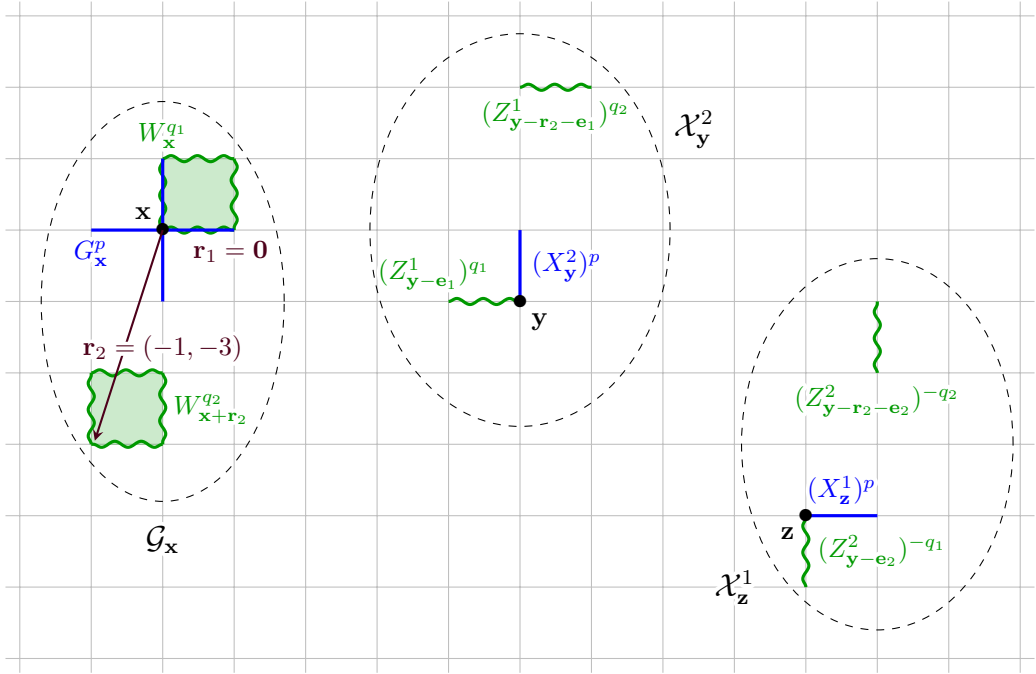
\begin{figure}[b!]
\begin{center}
\begin{tikzpicture}[scale = 0.95]
  \contourlength{1.5pt}

  \draw[step = 1, gray!50!white] (-0.2, -0.2) grid (14.2, 9.2);

  \fill[fill=green!60!black, fill opacity = 0.2] (1, 3) rectangle +(1, 1);
  \fill[fill=green!60!black, fill opacity = 0.2] (2, 6) rectangle +(1, 1);
  \draw[green!60!black, very thick,
    style = {decorate, decoration = {snake, amplitude = 0.4mm}}]
    (2, 6) -- (2, 7);
  \draw[green!60!black, very thick,
    style = {decorate, decoration = {snake, amplitude = 0.4mm}}]
    (2, 7) -- (3, 7);
  \draw[green!60!black, very thick,
    style = {decorate, decoration = {snake, amplitude = 0.4mm}}]
    (3, 7) -- (3, 6);
  \draw[green!60!black, very thick,
    style = {decorate, decoration = {snake, amplitude = 0.4mm}}]
    (2, 6) -- (3, 6);
  \draw[green!60!black, very thick,
    style = {decorate, decoration = {snake, amplitude = 0.4mm}}]
    (1, 3) -- (1, 4);
  \draw[green!60!black, very thick,
    style = {decorate, decoration = {snake, amplitude = 0.4mm}}]
    (1, 4) -- (2, 4);
  \draw[green!60!black, very thick,
    style = {decorate, decoration = {snake, amplitude = 0.4mm}}]
    (2, 4) -- (2, 3);
  \draw[green!60!black, very thick,
    style = {decorate, decoration = {snake, amplitude = 0.4mm}}]
    (2, 3) -- (1, 3);
  \draw[green!60!black, font = \footnotesize] (2, 7) node[above] {\contour{white}{$W_{\b x}^{q_1}$}};
  \draw[green!60!black, font = \footnotesize] (2, 3.5) node[right] {\contour{white}{$W_{\b x + \b r_2}^{q_2}$}};

  \draw[blue, very thick]
    (2, 5) -- (2, 7);
  \draw[blue, very thick]
    (1, 6) -- (3, 6);
  \draw[blue, font = \footnotesize] (1, 6) node[below] {\contour{white}{$G^p_{\b x}$}};

  \draw[thick] (2, 6) node {$\bullet$} node[above left, font = \footnotesize] {$\b x$};

  \draw[dashed] (2, 5) ellipse (1.7 and 2.8);
  \draw (2, 2) node[below] {\contour{white}{$\G_{\b x}$}};

  \draw[purple!40!black] (3 - 0.05, 6) node[below, font = \footnotesize] {\contour{white}{$\b r_1 = \b 0$}};
  \draw[purple!40!black, -stealth, thick] (2, 6) -- (1.05, 3.05);
  \draw[purple!40!black] (2, 4) node[above, font = \footnotesize]
  {\contour{white}{$\b r_2 = (-1, - 3)$}};

  \draw[blue, very thick] (7, 5) -- (7, 6) node[midway, right, font = \footnotesize] {\contour{white}{$(X_{\b y}^2)^p$}};
  \draw[green!60!black, very thick,
    style = {decorate, decoration = {snake, amplitude = 0.4mm}}]
    (6, 5) -- (7, 5) node[near end, above left, font = \footnotesize] {\contour{white}{$(Z_{\b y - \b e_1}^1)^{q_1}$}};
  \draw[green!60!black, very thick,
    style = {decorate, decoration = {snake, amplitude = 0.4mm}}]
    (7, 8) -- (8, 8) node[midway, below, font = \footnotesize] {\contour{white}{$(Z_{\b y - \b r_2 - \b e_1}^1)^{q_2}$}};

  \draw[thick] (7, 5) node {$\bullet$} node[below right, font = \footnotesize] {$\b y$};
  \draw[dashed] (7, 6) ellipse (2.1 and 2.75);
  \draw (9, 7) node[above right] {\contour{white}{$\X^2_{\b y}$}};

  \draw[blue, very thick] (11, 2) -- (12, 2) node[midway, above, font = \footnotesize] {\contour{white}{$(X_{\b z}^1)^p$}};
  \draw[green!60!black, very thick,
    style = {decorate, decoration = {snake, amplitude = 0.4mm}}]
    (11, 1) -- (11, 2) node[midway, right, font = \footnotesize] {\contour{white}{$(Z_{\b y - \b e_2}^2)^{-q_1}$}};
  \draw[green!60!black, very thick,
    style = {decorate, decoration = {snake, amplitude = 0.4mm}}]
    (12, 4) -- (12, 5) node[pos = 0, below, font = \footnotesize] {\contour{white}{$(Z_{\b y - \b r_2 - \b e_2}^2)^{-q_2}$}};

  \draw[thick] (11, 2) node {$\bullet$} node[below left, font = \footnotesize] {$\b z$};
  \draw[dashed] (12, 3) ellipse (1.9 and 2.6);
  \draw (10, 1) node {\contour{white}{$\X^1_{\b z}$}};

\end{tikzpicture}
\end{center}
\caption{\small An example of a generalized Gauss operator $\G_{\b x}$ with two attached magnetic fluxes, together with canonical gauge-invariant shift operators $\X_{\b y}^2$ and $\X_{\b z}^1$. The plaquettes hosting fluxes attached to site $\b x$ are shaded green. Each $\X_{\b x}^i$ consists of an elementary shift operator $(X_{\b x}^i)^p$, shown in blue, and of a product of elementary clock operators $Z_{\b y}^j$ over the chain $c(\b x, i)$ determined by the canonical solution \eqref{canonical flux attachment}, shown in green. This canonical choice has the fewest Wilson lines entering the $\X$'s. Other choices of $c(\b x, i)$ correspond to multiplying these $\X$'s by various Wilson loops.}
\label{fig flux attachment}
\end{figure}

The canonical solution \eqref{canonical flux attachment} is quite simple. For every plaquette $\b x + \b r_\alpha$ whose magnetic flux contributes to the generalized Gauss operator $\G_{\b x}$, the generalized shift operator $\X_{\b x}^i$ contains one Wilson line of charge $-q_\alpha$ that lives on the link perpendicular to the direction $i$ and terminates at point $\b x - \b r_\alpha$. This is depicted on Fig.\ \ref{fig flux attachment}.

\noindent \bt{A working example.} What rules of flux attachment --- i.e.\ what choices of $p$ and $\{(q_\alpha, \b r_\alpha)\}_\alpha$ --- lead to a fully confinable theory on a square lattice? The simplest examples are obtained by imposing the ``perturbative'' condition \eqref{conf condition pert} on one-chains given by \eqref{canonical flux attachment}. This results in
\bel{
  \sum_\alpha
    q_\alpha \epsilon^{ij} \delta_{\b x, \, \b y + \b r_\alpha + \b e_j}
   =
  \sum_\alpha
    q_\alpha \epsilon^{ji} \delta_{\b y, \, \b x + \b r_\alpha + \b e_i},
}
a condition that must hold for all $\b x, \b y \in \Mbb$ and $i, j \in \{1, 2\}$.

The na\"ive solution --- matching summands for each $\alpha$ --- results in the requirement
\bel{
  2\b r_\alpha = - \b e_1 - \b e_2 = (-1, -1).
}
Since any $\b r_\alpha$ must have integer entries, there are in fact no solutions to be found here.

\begin{figure}[b!]
\begin{center}
\begin{tikzpicture}[scale = 0.95]
  \contourlength{1.5pt}

  \draw[step = 1, gray!50!white] (-0.2, -0.2) grid (11.2, 6.2);

  \fill[fill=green!60!black, fill opacity = 0.2]  (2, 4) rectangle +(1, 1);
  \draw[green!60!black, very thick,
    style = {decorate, decoration = {snake, amplitude = 0.4mm}}]
    (2, 4) -- (2, 5);
  \draw[green!60!black, very thick,
    style = {decorate, decoration = {snake, amplitude = 0.4mm}}]
    (2, 5) -- (3, 5) node[midway, above, font = \footnotesize] {\contour{white}{$W_{\b x}^q$}};
  \draw[green!60!black, very thick,
    style = {decorate, decoration = {snake, amplitude = 0.4mm}}]
    (3, 5) -- (3, 4);
  \draw[green!60!black, very thick,
    style = {decorate, decoration = {snake, amplitude = 0.4mm}}]
    (3, 4) -- (2, 4);

  \fill[fill=green!60!black, fill opacity = 0.2]  (1, 3) rectangle +(1, 1);
  \draw[green!60!black, very thick,
    style = {decorate, decoration = {snake, amplitude = 0.4mm}}]
    (1, 3) -- (1, 4);
  \draw[green!60!black, very thick,
    style = {decorate, decoration = {snake, amplitude = 0.4mm}}]
    (1, 4) -- (2, 4);
  \draw[green!60!black, very thick,
    style = {decorate, decoration = {snake, amplitude = 0.4mm}}]
    (2, 4) -- (2, 3);
  \draw[green!60!black, very thick,
    style = {decorate, decoration = {snake, amplitude = 0.4mm}}]
    (2, 3) -- (1, 3) node[pos = 0.3, below, font = \footnotesize] {\contour{white}{$W_{\b x - \b e_1 - \b e_2}^{-q}$}};

  \draw[blue, very thick] (1, 4) -- (3, 4) node[near end, below, font = \footnotesize] {\contour{white}{$G_{\b x}^p$}};
  \draw[blue, very thick] (2, 3) -- (2, 5);

  \draw[thick] (2, 4) node {$\bullet$} node[above left, font = \footnotesize] {$\b x$};

  \draw[dashed] (2-0.1, 4-0.05) ellipse (1.75 and 2);
  \draw (2, 1.9) node[below] {\contour{white}{$\G_{\b x}$}};

  \draw[blue, very thick]
    (9, 4) -- (9, 5) node[midway, right, font = \footnotesize] {\contour{white}{$(X^2_{\b z})^{p}$}};
  \draw[green!60!black, very thick,
    style = {decorate, decoration = {snake, amplitude = 0.4mm}}]
    (8, 4) -- (9, 4) node[pos = 0, above, font = \footnotesize] {\contour{white}{$(Z_{\b x - \b e_1}^1)^{q}$}};
    \draw[green!60!black, very thick,
    style = {decorate, decoration = {snake, amplitude = 0.4mm}}]
    (9, 5) -- (10, 5) node[midway, above, font = \footnotesize] {\contour{white}{$(Z_{\b x + \b e_2}^1)^{-q}$}};
  \draw[thick] (9, 4) node {$\bullet$} node[below right, font = \footnotesize] {$\b z$};
  \draw[dashed] (9, 4.6) ellipse (2 and 1.5);
  \draw (7, 5) node[above left] {$\X_{\b z}^2$};

  \draw[green!60!black, very thick,
    style = {decorate, decoration = {snake, amplitude = 0.4mm}}]
    (5, 2) -- (5, 1) node[midway, right, font = \footnotesize] {\contour{white}{$(Z_{\b x - \b e_2}^2)^{-q}$}};
  \draw[green!60!black, very thick,
    style = {decorate, decoration = {snake, amplitude = 0.4mm}}]
    (6, 2) -- (6, 3) node[midway, right, font = \footnotesize] {\contour{white}{$(Z_{\b x + \b e_1}^2)^{q}$}};
  \draw[blue, very thick]
    (5, 2) -- (6, 2) node[near start, above, font = \footnotesize] {\contour{white}{$(X_{\b x}^1)^{p}$}};
  \draw[thick] (5, 2) node {$\bullet$} node[left, font = \footnotesize] {\contour{white}{$\b y$}};
  \draw[dashed] (6.1, 2) ellipse (2 and 1.5);
  \draw (7.5, 1) node[below right] {\contour{white}{$\X_{\b y}^1$}};

\end{tikzpicture}
\end{center}
\caption{\small Operators $\G_{\b x}$, $\X_{\b y}^1$ and $\X_{\b z}^2$ in a theory with flux attachment parameters given by \eqref{double flux attachment}.}
\label{fig double flux attachment}
\end{figure}
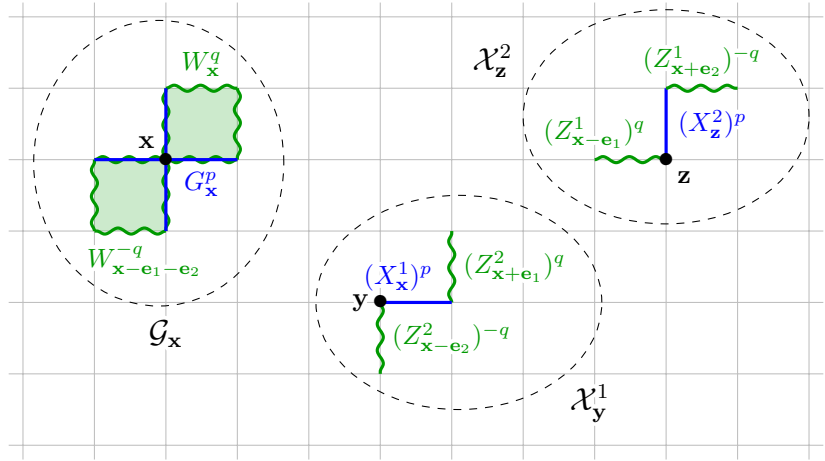

The next option is to have pairs of $\alpha$'s match up. This means that, for every $\alpha$, there exists a $\beta$ such that
\bel{
  q_\alpha \epsilon^{ij} \delta_{\b x, \, \b y + \b r_\alpha + \b e_j}
   =
  q_\beta \epsilon^{ji} \delta_{\b y, \, \b x + \b r_\beta + \b e_i}.
}
This translates to
\bel{
  q_\alpha + q_\beta = 0, \quad \b r_\alpha + \b r_\beta = - \b e_1 - \b e_2.
}
These conditions \emph{can} be solved. Possibly the simplest solution has two attached fluxes with
\bel{\label{double flux attachment}
  q_1 = - q_2 = q, \quad \b r_1 = \b 0, \quad \b r_2 = - \b e_1 - \b e_2.
}
As shown on Fig.\ \ref{fig double flux attachment}, the corresponding theory has two opposite magnetic fluxes positioned symmetrically across the site hosting the elementary Gauss operator.

\noindent \bt{The action.} The rest of this Section will focus on the fully confined phase of the theory with flux attachment parameters \eqref{double flux attachment}. Consider, first, the action that describes this phase. It can be obtained from the general expression \eqref{def S1 CS full conf} by a straightforward set of substitutions:
\begin{itemize}
  \item The sum over links $\ell$ is written as a sum over sites $\b x \in \Mbb$ and directions $i \in \{1, 2\}$.
  \item The coboundary operator acting on $A^0_{\b x}$ gives $(\delta A^0)_{\b x}^i = (\del_i A^0)_\b x \equiv A^0_{\b x + \b e_i} - A^0_{\b x}$.
  \item The combination $(\del_0 A^i)_{\b x} - (\del_i A^0)_{\b x}$ is denoted $F^{0i}_{\b x}$, anticipating the introduction of the field strength tensor.
  \item The one-form $A_{c(\ell)}$ at $\ell \equiv (\b x, i)$ becomes $A_{c(\b x, i)} \equiv q \sum_j \epsilon^{ij} (A^j_{\b x + \b e_i} - A^j_{\b x - \b e_j})$.
  \item The two-form $B_{a(v)}$ at $v \equiv \b x$ becomes $q(B_{\b x} - B_{\b x - \b e_1 - \b e_2})$.
  \item Finally, $B_{\b x}$ is denoted $F^{12}_{\b x}$, and the remaining components of the field strength are defined via $F^{\mu \nu}_{\b x} = -F^{\nu\mu}_{\b x}$.
\end{itemize}

Doing all this results in the one-derivative action
\bel{\label{def S1 double confinement}
  S_1[A, A^0]
   =
  -\frac{qK}{2\pi p}
  \sum_{\tau \in \Sbb} \d\tau \sum_{\b x \in \Mbb} \left(\L_{\b x, \tau}^+ - \L_{\b x, \tau}^- \right)
}
with
\gathl{
  \L_{\b x, \tau}^+
   \equiv
  A^2_{\b x + \b e_1, \tau} F^{01}_{\b x, \tau}
   +
  A^1_{\b x + \b e_2, \tau} F^{20}_{\b x, \tau}
   +
  A_{\b x, \tau}^0 F^{12}_{\b x, \tau + \d\tau},
   \\
  \L_{\b x, \tau}^-
   \equiv
  A^2_{\b x - \b e_2, \tau} F^{01}_{\b x, \tau}
   +
  A^1_{\b x - \b e_1, \tau} F^{20}_{\b x, \tau}
   +
  A^0_{\b x, \tau} F^{12}_{\b x - \b e_1 - \b e_2, \tau + \d\tau}.
}
These Lagrangians are examples of discretizations of the continuum CS term $\epsilon^{\mu\nu\lambda} A\_c^\mu \del_\nu A\_c^\lambda$. In particular, choosing $q = -1$ and $p = 2$ makes the corresponding actions $S_1^\pm$ have precisely the normalization $\pm K/4\pi$, as appropriate for a U(1) CS action at level $\pm K$.

While the parallels with continuum CS are striking, the differences must not be ignored:
\begin{itemize}
  \item There are two CS terms of opposite levels. This situation is reminiscent of fermion doubling. However, here the ``doubling'' only appears at the level of the action and is invisible in the canonical formalism. Moreover, there is no way to decouple the ``doublers'': the same field variables will appear in both actions. This means that $S_1$ is, in effect, a two-derivative action.
  \item The integration variables --- in particular the fields $A^i_{\b x}$ --- take values in a finite set. This is especially important when $K$ is very small. It is impossible to set $p = 2$ in a theory with $K = 2$. Even worse, there is no meaningful lattice theory of the above form that gives rise to the invertible U(1)$_1$ CS theory. The above construction only resembles two pairs of continuum CS actions when $K \gg 1$.
\end{itemize}

\noindent \bt{Ground state degeneracy.} The fully confined theory defined by \eqref{double flux attachment} does not show CS behavior at low energies. Nevertheless, this theory has some interesting features on its own. A particularly noteworthy one is that its ground state degeneracy delicately depends on the dimensions of the lattice. This effect is reminiscent of fracton theories \cite{Nandkishore:2018sel}, but here there are no particles of limited mobility. The nontrivial degeneracy is ultimately caused by the fact that gauge constraints with $p \neq 1$ do not affect certain topological $\Z_p$ degrees of freedom.

Here is how the counting of states in the confined sector works when $q = 1$ and $p | K$. Before any constraints are imposed, there are $K$ states available on each link, for a total of $K^{|\Mbb|_1} = K^{2N_1 N_2}$ linearly independent states in the Hilbert space. Na\"ively, it may seem that setting $\X_{\b x}^i = \1$ imposes $K$ linearly independent constraints at each link --- one for each nontrivial power of $\X_{\b x}^i$ --- and hence there are a total of $K^{2N_1N_2}$ independent constraints that single out a unique ground state. However, there are actually relations between some of the constraints on different links. To see this, consider the product of operators $\X_{\b x}^i$ over a diagonal, i.e.\ over all $\b x$ of the form $\b x + n(\b e_1 + \b e_2)$ for integer $n$, modulo the periodicity condition $\b x \equiv \b x + N_i \b e_i$. All the Wilson lines in this product cancel, and what remains is just the product of operators $(X^i_{\b x})^p$ over the diagonal. This product becomes the identity when raised to the $K/p$'th power. Thus the actual number of linearly independent constraints is $K^{2N_1N_2}/p^{2N\_{diag}}$, where $N\_{diag}$ is the number of distinct diagonals in the system. If $N_1 = N_2 \equiv N$, this number is $N\_{diag} = N$, and the ground state degeneracy is $p^{2N}$. If $N_1$ and $N_2$ are coprime, the number of diagonals is $N\_{diag} = 1$, and the ground state degeneracy is $p^2$. In general, of course, this theory has
\bel{
  p^{2\,\trm{gcd}(N_1, N_2)}
}
ground states. This number generically grows exponentially with the \emph{linear} size of the system. The ground states are collective $\Z_p$ modes of electric flux-eigenstates along all links piercing a diagonal, with the direction of the diagonal determined by the rules of flux attachment.

\noindent \bt{Entanglement entropies.} Let $S_\Vbb$ be the entanglement entropy associated to a rectangular region $\Vbb$ with $M_1 \times M_2$ plaquettes. A standard result in ordinary $\Z_K$ gauge theories is that
\bel{
  S_\Vbb\^{(ordinary)} =
  \left\{
    \begin{array}{ll}
      0 & \hbox{in the confined phase,} \\
      \left(|\del\Vbb| - 1\right) \log K & \hbox{in the topological phase,}
    \end{array}
  \right.
}
where the length of the entangling edge is $|\del \Vbb| = 2(M_1 + M_2)$. (See \cite{Lin:2018bud} for a review.) A fully confined state of the $q = 1$ flux-attached theory \eqref{def S1 double confinement} has $M_1 (M_2 - 1) + M_2 (M_1 - 1)$ linearly independent $\Z_K$ constraints imposed upon the $M_1 (M_2 + 1) + M_2(M_1 + 1)$ degrees of freedom in $\Vbb$. This means that the reduced density matrix associated to $\Vbb$ is proportional to the identity matrix of size $K^{2(M_1 + M_2)}$. In terms of the size of $\Vbb$, its von Neumann entropy is
\bel{\label{EE}
  S_\Vbb = |\del \Vbb| \log K.
}

At first glance, the absence of a $|\del \Vbb|$-independent term in \eqref{EE} suggests that there is no long-range entanglement in the ground state of a fully confined flux-attached theory. However, consider modifying the region $\Vbb$ by removing the two links emerging from the upper left corner of $\Vbb$.  The new region $\Vbb'$ has an entangling edge of the same size as before, but the entropy is
\bel{
  S_{\Vbb'} = \left(|\del \Vbb| - 2\right) \log K.
}
This shows that entanglement entropy in the fully confined phase is sensitive to the microscopic properties of the entangling edge, in contrast to the entanglement entropy in the topological phase. It is thus not possible to decide whether the confined phase has long-range entanglement by looking at the entropy associated to just one region.

\begin{figure}
\begin{center}
\begin{tikzpicture}[scale = 0.95]
  \contourlength{0.3pt}

  \draw[step = 0.2, gray!30!white] (-0.3, -0.1) grid (5.1, 3.7);

  \fill[fill=green!60!black, fill opacity = 0.5]  (1, 1.6) rectangle +(1.4, 1) node[pos = 0.5, green] {\contour{green!20!black}{$\Vbb_1$}};
  \fill[fill=red!60, fill opacity = 0.5]  (2.4, 1.6) rectangle +(1.4, 1)
  node[pos = 0.5, red!50!black] {\contour{red!70!black}{$\Vbb_2$}};
  \fill[fill=blue!60, fill opacity = 0.5]  (1, 0.6) rectangle +(2.8, 1)
  node[pos = 0.5, blue!50!black] {\contour{blue!70!black}{$\Vbb_3$}};

  \draw[<->, thick] (1, 2.75) -- (2.4, 2.75) node[midway, above] {$M_1$};
  \draw[<->, thick] (2.4, 2.75) -- (3.8, 2.75) node[midway, above] {$M_1$};
  \draw[<->, thick] (0.85, 0.6) -- (0.85, 1.6) node[midway, left] {$M_2$};
  \draw[<->, thick] (0.85, 1.6) -- (0.85, 2.6) node[midway, left] {$M_2$};

\end{tikzpicture}
\end{center}
\caption{\small A choice of regions used to compute the topological entanglement entropy.}
\label{fig TEE}
\end{figure}
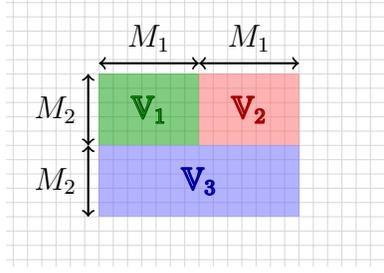

A more reliable way to detect long-range entanglement is by computing the topological entanglement entropy (TEE) \cite{Kitaev:2005dm, Levin:2006}. Consider the three regions $\Vbb_1$, $\Vbb_2$, and $\Vbb_3$ shown on Fig.\ \ref{fig TEE}. By the Kitaev-Preskill prescription \cite{Kitaev:2005dm}, the TEE is given by the tripartite information
\bel{
  S\_{topo}\^{KP} = S_{\Vbb_1} + S_{\Vbb_2} + S_{\Vbb_3} - S_{\Vbb_1 \cup \Vbb_2} - S_{\Vbb_2 \cup \Vbb_3} - S_{\Vbb_1 \cup \Vbb_3} + S_{\Vbb_1 \cup \Vbb_2 \cup \Vbb_3}.
}
The entropies of these regions are all readily evaluated, and they are
\gathl{
  S_{\Vbb_1} = S_{\Vbb_2} = 2(M_1 + M_2) \log K,
   \quad
  S_{\Vbb_1 \cup \Vbb_2} = S_{\Vbb_3} = 2(2M_1 + M_2) \log K, \\
  S_{\Vbb_1 \cup \Vbb_3} = S_{\Vbb_1 \cup \Vbb_2 \cup \Vbb_3} = 4(M_1 + M_2) \log K,
   \quad
  S_{\Vbb_2 \cup \Vbb_3} = \left[4(M_1 + M_2) - 2\right] \log K.
}
It follows that the TEE is
\bel{
  S\_{topo}\^{KP} = 2\log K.
}
The fact that $S\_{topo}\^{KP} \neq 0$  suggests that nontrivial long-range entanglement is present.

A remarkable fact about this result is that the TEE is \emph{positive}. This sets the theory \eqref{def S1 double confinement} apart from all conventional topological QFTs, in which $\D \equiv \e^{-S\^{KP}\_{topo}}$, the \emph{total quantum dimension}, is a universal number bounded from below by unity. In the flux-attached theory at hand, the total quantum dimension appears to be $\D = 1/K^2 < 1$.

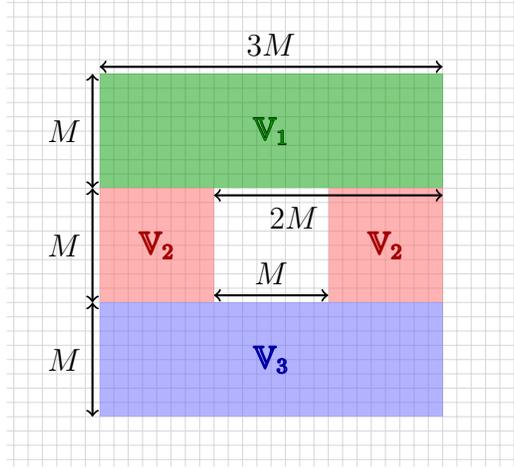
\begin{figure}
\begin{center}
\begin{tikzpicture}[scale = 0.95]
  \contourlength{0.3pt}

  \draw[step = 0.2, gray!30!white] (-0.3, -0.1) grid (6.9, 6.5);

  \fill[fill=green!60!black, fill opacity = 0.5]  (1, 3.8) rectangle +(4.8, 1.6) node[pos = 0.5, green] {\contour{green!20!black}{$\Vbb_1$}};
  \fill[fill=red!60, fill opacity = 0.5]  (1, 2.2) rectangle +(1.6, 1.6)
  node[pos = 0.5, red!50!black] {\contour{red!70!black}{$\Vbb_2$}};
  \fill[fill=red!60, fill opacity = 0.5]  (4.2, 2.2) rectangle +(1.6, 1.6)
  node[pos = 0.5, red!50!black] {\contour{red!70!black}{$\Vbb_2$}};
  \fill[fill=blue!60, fill opacity = 0.5]  (1, 0.6) rectangle +(4.8, 1.6)
  node[pos = 0.5, blue!50!black] {\contour{blue!70!black}{$\Vbb_3$}};

  \draw[<->, thick] (1, 5.5) -- +(4.8, 0) node[midway, above] {$3M$};
  \draw[<->, thick] (2.6, 2.3) -- +(1.6, 0) node[midway, above] {$M$};
  \draw[<->, thick] (2.6, 3.7) -- +(3.2, 0) node[pos = 0.35, below] {$2M$};
  \draw[<->, thick] (0.9, 0.6) -- +(0, 1.6) node[midway, left] {$M$};
  \draw[<->, thick] (0.9, 2.2) -- +(0, 1.6) node[midway, left] {$M$};
  \draw[<->, thick] (0.9, 3.8) -- +(0, 1.6) node[midway, left] {$M$};

\end{tikzpicture}
\end{center}
\caption{\small Another choice of regions used to compute the topological entanglement entropy.}
\label{fig TEE LW}
\end{figure}

This unconventional result should give one pause. A further surprise comes from employing the \emph{other} standard method for computing the TEE, due to Levin and Wen \cite{Levin:2006}. Consider once again three regions, this time arranged as shown on Fig.\ \ref{fig TEE LW}, and once again compute
\bel{
  S\_{topo}\^{LW} = S_{\Vbb_1} + S_{\Vbb_2} + S_{\Vbb_3} - S_{\Vbb_1 \cup \Vbb_2} - S_{\Vbb_2 \cup \Vbb_3} - S_{\Vbb_1 \cup \Vbb_3} + S_{\Vbb_1 \cup \Vbb_2 \cup \Vbb_3}.
}
The individual entropies are
\gathl{
  S_{\Vbb_1} = S_{\Vbb_2} = S_{\Vbb_3} = 8M \log K, \quad
  S_{\Vbb_1 \cup \Vbb_2} = S_{\Vbb_2 \cup \Vbb_3} = (12M - 2) \log K, \\
  S_{\Vbb_1 \cup \Vbb_3} = 16 M \log K, \quad
  S_{\Vbb_1 \cup \Vbb_2 \cup \Vbb_3} = (16M - 4) \log K.
}
The Levin-Wen TEE thus vanishes,
\bel{
  S\_{topo}\^{LW} = 0,
}
instead of being equal to $2S\_{topo}\^{KP}$, as in conventional topological QFTs. This should be taken as a strong indication that the fully confined theory is not topologically invariant.

This pathological behavior of the TEE is not completely new. It has recently been argued that the standard protocols for calculating the TEE receive spurious contributions if the underlying theory possesses rigid subsystem symmetries \cite{Cano:2014pya, Zou:2016dck, Santos:2018eai, Williamson:2018zig}. Indeed, there is a sense in which the present theory features special subsystems: the fully confined state entangles only links that intersect the same SW-NE diagonal. Therefore each entanglement entropy can be decomposed as $S_\Vbb = \sum_{\delta \subset \Vbb} S_\delta$, with each $\delta$ being a subset of links of $\Vbb$ that intersect the same diagonal, and so the state can be realized as a stack of unentangled one-dimensional quantum chains. However, it should be noted that this is a special feature of the choice \eqref{double flux attachment}. Confinement in other flux-attached theories may yet give conventional topological orders.

%\noindent \bt{Duality.} Another interesting perspective on this theory comes from considering its dual forms. While the original flux-attached theory, with a single $\b r_\alpha = \b 0$, is dual to $\Z_K$ parafermions, the theory \eqref{double flux attachment} actually dualizes to \emph{bosons} for any $K$. Specifically, consider a system of $\Z_K$ quantum rotors on plaquettes of $\Mbb$ (or, equivalently, on sites of the dual lattice $\Mbb^\vee$).

\newpage

\section{A partially confinable $\Z_2$ theory} \label{sec part conf}

\bt{The model.} This Section will stay on a square lattice, but will specialize to the case $K = 2$. In this situation there is only one nontrivial value of the parameters $p$ and $q_\alpha$ in a generalized Gauss operator \eqref{def G general}: unity. The only choice involved in the rules of flux attachment thus lies in the freedom to pick different values of $\b r_\alpha$.

When these rules are chosen so that the theory is only partially confinable, one way to proceed is to impose as many confining constraints as possible and obtain a low energy theory with a finite-dimensional gauge-invariant Hilbert space. One such partially confined theory, with $\X_{\b x}^2 = \1$ imposed at every site $\b x$, was shown to lead to a rather strange kind of CS-like theory that retains a delicate sensitivity to the parity $N\, \trm{mod}\, 2$ of the linear size of the lattice \cite{Radicevic:3D}. This kind of theory does not have a conventional continuum limit.

However, when $K = 2$, it is possible to precisely study a more reasonable partially confinable theory: the $g \rar \infty$ limit of the Kogut-Susskind Hamiltonian \eqref{def H KS} with $\b r_\alpha = \b 0$, given by
\bel{\label{def H KS strong}
  H_{\infty} = - \sum_{\b x, i} \X_{\b x}^i
}
after a shift and a rescaling. This theory is \emph{not} partially confined. In fact, $H_\infty$ is not a sum of commuting projectors, so it is not possible to study it just by counting constraints \`a la Section \ref{sec square}. The way to study it, instead, is by exploiting the property of $\Z_2$ flux-attached theories that lead to their original introduction: the fact that they are dual to fermions \cite{Chen:2017fvr}.

\noindent \bt{Bosonization.} Recall that there exists a higher-dimensional Jordan-Wigner transformation that can be applied to a $\Z_2$ gauge theory with generalized Gauss operators $\G_{\b x} = G_{\b x} W_{\b x}$. This is an exact duality that maps all local gauge-invariant operators to local bilinears of spinless fermions that live on sites of the dual lattice. The dual lattice will not be explicitly introduced in this paper, and the dual fermions will simply be regarded as living on plaquettes of the original square lattice $\Mbb$. With this convention, the basic form of the duality is
\bel{\label{JW basic}
  \X_{\b x}^1 = - \i \chi'_{\b x} \chi_{\b x - \b e_2},  \quad
  \X_{\b x}^2 = - \i \chi'_{\b x - \b e_1} \chi_{\b x},  \quad
  W_{\b x} = \i \chi'_{\b x} \chi_{\b x},
}
where $\chi_{\b x}$ and $\chi_{\b x}'$ are Majorana operators associated to the plaquette labeled by $\b x$. In terms of the more familiar complex fermion operators on $\b x$, these Majoranas are
\bel{\label{def psi}
  \chi_{\b x} = \psi_{\b x}\+ + \psi_{\b x}, \quad
  \chi'_{\b x} = \i \left(\psi_{\b x}\+ - \psi_{\b x} \right),
}
and their commutation relation is
\bel{
  \{\chi_{\b x}, \chi_{\b y}\} = \{\chi_{\b x}', \chi_{\b y}'\} = 2\delta_{\b x, \b y}\, \1, \quad
  \{\chi_{\b x}, \chi_{\b y}'\} = 0.
}

A subtle point is that \eqref{JW basic} is a ``singlet-singlet'' duality on a toric lattice \cite{Radicevic:2018okd}. This means that it does not map all states in the gauge theory to all states in the fermion theory. Instead, it maps the singlet sector of the fermion parity $(-1)^F \equiv \prod_{\b x} \i \chi'_{\b x} \chi_{\b x}$ to the singlet sector of the one-form symmetry generated by line operators of the form $\prod_{x_2} X_{\b x}^1 Z_{\b x + \b e_1}^2$.\footnote{
  These operators can be viewed as products of Wilson and 't Hooft lines that wind along the same noncontractible cycle. In the toric code language, they are lines associated to the $\epsilon$ particle.}
This is easily seen by multiplying both sides of the duality relation $W_{\b x} = \i \chi_{\b x}' \chi_{\b x}$ over all points $\b x \in \Mbb$, or by multiplying e.g.\ $\chi'_{\b x} \chi'_{\b x - \b e_2} = W_{\b x - \b e_2} \X^1_{\b x}$ over all $\b x$ with a fixed value of $x_1$. This means that the duality \eqref{JW basic} does not map all states of the flux-attached gauge theory, and so it is not possible to use it to find a \emph{complete} solution to the gauge theory \eqref{def H KS strong}.

It is possible to modify the basic Jordan-Wigner duality by ``twisting'' it, i.e.\ by introducing flat background $\Z_2$ gauge fields for the $(-1)^F$ symmetry on the fermionic side \cite{Radicevic:2018okd}. Then the Wilson lines of these fields along noncontractible cycles dualize to the nontrivial one-form symmetry generators in the flux-attached gauge theory. The upshot is that the duality that maps all gauge-invariant states in the theory \eqref{def H KS strong} has the form
\bel{\label{JW twisted}
  \X_{\b x}^1 = - \i \chi'_{\b x} \zeta^2_{\b x - \b e_2} \chi_{\b x - \b e_2},  \quad
  \X_{\b x}^2 = - \i \chi'_{\b x - \b e_1} \zeta^1_{\b x - \b e_1} \chi_{\b x},  \quad
  W_{\b x} = \i \chi'_{\b x} \chi_{\b x},
}
where $\zeta^i_{\b x}$ are c-numbers that take values $\pm 1$ and satisfy
\bel{
  \zeta^1_{\b x}  \zeta^2_{\b x + \b e_1} \zeta^1_{\b x + \b e_2} \zeta^2_{\b x} = 1 \quad \trm{for\ each\ }\b x \in \Mbb.
}
This is the flatness constraint for $\Z_2$ gauge fields that live on links of the dual lattice.

The map \eqref{JW twisted} underdetermines the gauge fields $\zeta^i_{\b x}$. For example, given one flat configuration, say $\zeta^i_{\b x} = 1$ for every link $(\b x, i)$, it is possible to generate many other flat configurations by performing ``gauge transformations,'' i.e.\ by flipping sets of four $\zeta$'s that emanate from the same site. Each of these configurations leads to a \emph{different} exact duality, i.e.\ to a different way of rewriting the flux-attached theory \eqref{def H KS strong} in terms of fermions. However, every ``gauge transformation'' of this form corresponds to conjugating the Hamiltonian by one of the $W_{\b x}$ operators, and so all these dual theories have the same energy spectrum. For the purposes of solving \eqref{def H KS strong}, it is thus sufficient to pick the most convenient configuration of $\zeta$'s from the orbit of local ``gauge transformations.''

The only truly inequivalent configurations are those related by nontrivial 't Hooft loops, i.e.\ the ones related to each other by flipping all $\zeta^i_{\b x}$'s at a single fixed $x^i$, say $x^i = N$. Each of these four distinct configurations captures a different superselection sector of the one-form symmetry of the flux-attached theory. In the fermionic theory, these correspond to different choices of boundary conditions for fermions. In other words, one must sum over all spin structures in the fermion theory to fully match the physics of the flux-attached theory.

\noindent \bt{The solution.} In order to solve the theory \eqref{def H KS strong} it is thus sufficient to study the quadratic fermion model
\bel{\label{def H F}
  H\_F = \i \sum_{\b x \in \Mbb} \left[\chi_{\b x}' \chi_{\b x + \b e_1} + \chi_{\b x + \b e_2}' \chi_{\b x} \right]
}
for all possible choices of (anti)periodic boundary conditions for the fermions. (This corresponds to choosing $\zeta^i_{\b x} = 1$ everywhere except on the ``boundary'' links.) It is convenient to first revert to complex fermions via \eqref{def psi}, getting
\bel{
  H\_F
   =
  -\sum_{\b x \in \Mbb} \left[
    \psi_{\b x}\+ \psi_{\b x + \b e_1} + \psi_{\b x}\+ \psi_{\b x + \b e_2} + \psi_{\b x}\+ \psi_{\b x + \b e_1}\+ +\psi_{\b x + \b e_2}\+ \psi_{\b x}\+
    + \trm{H.c.}
  \right].
}
Now introduce the Fourier transform
\bel{
  \psi_{\b x}
   \equiv
  \frac 1N \sum_{\b k \in \Pbb}
    \psi_{\b k} \, \e^{\frac{2\pi \i} N (\b k + \b s) \b x},
   \quad
  \Pbb \equiv \{\b k\}_{-\frac N2 \leq k^i < \frac N2},
}
where $\b s \in \{\b 0, \frac12 \b e_1, \frac12 \b e_2, \frac12 (\b e_1 + \b e_2)\}$ denotes the four possible spin structures, and where it is assumed (for convenience) that the lattice dimensions $N_1$ and $N_2$ are equal to an even number $N$. This leads to the Hamiltonian
\bel{
  H\_F
   =
  -\sum_{\b k \in \Pbb} \left[
    A_{\b k + \b s} \, \psi_{\b k}\+ \psi_{\b k}
     +
    B_{\b k + \b s} \, \psi_{\b k}\+ \psi_{-\b k - 2\b s}\+
     +
    \trm{H.c.}
  \right],
}
with
\bel{
  A_{\b k}
   \equiv
  \e^{\frac{2\pi\i}N k^1} + \e^{\frac{2\pi\i}N k^2},
   \quad
  B_{\b k}
   \equiv
  \e^{\frac{2\pi\i}N k^1} + \e^{-\frac{2\pi\i}N k^2}.
}

The standard way to diagonalize this Hamiltonian is by performing a Bogolyubov transformation. First, let
\bel{
  \Psi_{\b k} = \bcol{\Psi_{\b k}^+}{\Psi_{\b k}^-} \equiv \bcol{\psi_{\b k}}{\psi_{- \b k - 2\b s}\+},
}
Now, for a given $\b s$, it is natural to define two disjont momentum subspaces, $\Pbb^\pm(\b s)$, such that for every $\b k \in \Pbb^+(\b s)$ its ``reflection'' $-\b k - 2\b s$ belongs to $\Pbb^-(\b s)$. When $\b s \neq \b 0$, each $\b k \in \Pbb$ can be assigned to one of the two subsets. However, when $\b s = \b 0$, the momenta in $\Pbb^0(\b 0) \equiv \{\b 0, -\frac N2 \b e_1, -\frac N2 \b e_2, -\frac N2 (\b e_1 + \b e_2)\}$ do not belong to either $\Pbb^+(\b 0)$ or $\Pbb^-(\b 0)$. In other words, one can write
\bel{
  \Pbb = \Pbb^+(\b s) \cup \Pbb^-(\b s) \cup \Pbb^0(\b s),
}
with $\Pbb^0(\b s) = \varnothing$ if $\b s \neq \b 0$. With this convention, for every $\b k \in \Pbb^\pm(\b s)$ there exists the useful relation
\bel{\label{Psi reflection}
  \Psi_{- \b k - 2\b s}^\alpha = (\Psi_{\b k}^{-\alpha})\+.
}

Taking into account \eqref{Psi reflection} allows the Hamiltonian to be written as a two-band system
\gathl{
  H\_F
%   =
%  -\!\!\!\sum_{\b k \in \Pbb^+(\b s)} \!\!\! \left[
%    A_{\b k + \b s} (\Psi_{\b k}^+)\+ \Psi^+_{\b k} +
%    A_{\b k + \b s}^* \Psi_{\b k}^- (\Psi_{\b k}^-)\+ +
%    B_{\b k + \b s} (\Psi_{\b k}^+)\+ \Psi^-_{\b k} +
%    B_{\b k + \b s}^* \Psi_{\b k}^- (\Psi_{\b k}^+)\+ +
%    \trm{H.c.}
%  \right] \\
   =
  - \sum_{\b k \in \Pbb^+(\b s)}
    (\Psi^\alpha_{\b k})\+ D_{\b k + \b s}^{\alpha\beta} \Psi^\beta_{\b k}
  + H_0(\b s).
}
Here the ``band matrix'' is given by
\bel{\label{def D}
  D_{\b k}
   \equiv
  \bmat
   {A_{\b k} + A^*_{\b k}}
   {B_{\b k} - B^*_{\b k}}
   {B_{\b k}^* - B_{\b k}}
   {-A_{\b k} - A^*_{\b k}}.
}
Its eigenvalues --- i.e.\ the dispersions of excitations associated to a given momentum --- are
\bel{
  \pm\eps_{\b k}
   \equiv
  \pm \sqrt{|A_{\b k} + A^*_{\b k}|^2 + |B_{\b k} - B^*_{\b k}|^2}
  % =
  %\pm 2\sqrt{ \left(\cos \tfrac{2\pi}N k^1 + \cos \tfrac{2\pi}N k^2\right)^2
  %  + \left(\sin \tfrac{2\pi}N k^1 - \sin \tfrac{2\pi}N k^2 \right)^2 }
   %=
  %\pm 2\sqrt{2 + 2\cos \tfrac{2\pi}N (k^1 + k^2)}
   =
  \pm 4 \left|\cos \tfrac \pi N (k^1 + k^2) \right|.
}
Meanwhile, $H_0(\b s)$ collects the various terms proportional to the identity, and also those that correspond to momenta in $\Pbb^0(\b s)$. Using the facts that $\sum_{\b k \in \Pbb^0(\b 0)} A_{\b k} = 0$ and $\sum_{\b k \in \Pbb} A_{\b k} = 0$, this term can be simply written as
\bel{
  H_0(\b s)
   =
  4 \delta_{\b s, \b 0} \left(
     \psi_{-\frac N2(\b e_1 + \b e_2)}\+ \psi_{-\frac N2(\b e_1 + \b e_2)}
     - \psi_{\b 0}\+ \psi_{\b 0}
  \right).
}

This is all the information needed to obtain the full spectrum of the theory \eqref{def H KS strong}. Recall that there are four superselection sectors, each corresponding to a different choice of spin structure $\b s$. In each sector, the ground state subspace is characterized by all negative-dispersion modes being filled, and all positive-dispersion modes being empty. The overall ground state degeneracy is given by the number of zero-dispersion modes in the sector(s) with the lowest vacuum energy.\footnote{
  Note that there is a subtlety here: only $(-1)^F = 1$ states in the fermion theory have duals in the flux-attached theory. This is why the ground state degeneracy in each superselection sector equals the number of zero-dispersion modes and not the number of \emph{states} associated with zero-dispersion modes, as the latter number includes both fermion-odd and fermion-even states.}
By judiciously picking sets $\Pbb^+(\b s)$, the vacuum energy  corresponding to each sector (obtained by summing over negative-energy modes $-\eps_{\b k + \b s}$) can be expressed as
\bel{
  \E_0(\b s)
   =
  - 4 \sum_{k^1 = -\frac N2}^{\frac N2 - 1} \sum_{k^2 = 0}^{\frac N2 - 1}
    \left|\cos\frac\pi N \left(k^1 + k^2 + s^1 + s^2\right) \right|.
}
This quantity only depends on whether $s^1 + s^2$ is integer or half-integer. The sectors with $\b s = \b 0$ and $\b s = \frac 12 (\b e_1 + \b e_2)$ both have $\E_0(\b s) = -2N \cot\frac\pi{2N}$, while the other two sectors with $\b s = \frac 12 \b e_i$ have $\E_0(\b s) = - 2N \csc \frac{\pi}{2N} < - 2N \cot \frac{\pi}{2N}$. (It is not an accident that the same vacuum energies are found for the two sectors of the Ising model in $(1+1)$D, cf.\ \cite{McCoy:2014, Radicevic:2019mle}.) The global ground state subspace is thus spanned by the ground states of the latter two sectors.

Each superselection sector features a Fermi surface defined by $k^1 + k^2 + s^1 + s^2 = \frac N2 \, \trm{mod}\, N$. This is a generic phenomenon for bilinear fermion Hamiltonians that involve only local terms. A similar Fermi surface appears, for example, in the na\"ive Dirac Hamiltonian $\i \sum_{\b x, i} \left[\psi_{\b x}\+ \psi_{\b x + \b e_i} - \psi_{\b x + \b e_i}\+ \psi_{\b x} \right]$.

This surface is not stable under perturbations. One way to remove it is to include the first $1/g^2$ correction to the Hamiltonian \eqref{def H KS strong}. This amounts to adding a term of the form $-\lambda W_{\b x} = - \i \lambda \chi'_{\b x} \chi_{\b x}$, $\lambda > 0$, to the Hamiltonian density in \eqref{def H F}. This shifts the parameter $A_{\b k}$ in the dispersion relation, so the eigenvalues of the new band matrix are
\bel{\label{def eps(k) lambda}
  \pm \eps_{\b k}
   =
  \pm 2\sqrt{
    \lambda^2
    + 4\lambda \cos\frac\pi N (k^1 + k^2) \cos\frac\pi N (k^1 - k^2)
    + 4 \cos^2\frac\pi N (k^1 + k^2)
  }.
}
When $0 < \lambda < 2$, the function $\eps_{\b k}$ only vanishes at $k^1 = k^2 = \pm \frac{N}{2\pi} \arccos(-\frac\lambda 2)$. At $\lambda = 2$ these two points merge to give a single Dirac cone at $k^1 = k^2 = N/2$. The resulting critical theory lies at the transition between strong and weak coupling phases of the flux-attached Kogut-Susskind Hamiltonian \cite{Chen:2017fvr}.

There are other ways to remove the Fermi surface. One option it is to modulate the hopping terms by introducing the so-called Kawamoto-Smit signs \cite{Kawamoto:1981hw}. A judicious choice of these signs may replace the Fermi surface by a discrete set of Dirac cones. If such a choice exists, it would be further possible to gap such a modulated theory by introducing mass terms. These options will be studied elsewhere.

\noindent \bt{Topological degeneracy.} The low-energy behavior of \eqref{def H KS strong} displays all the usual degeneracies and singularities associated to the presence of a Fermi surface. However, there is an additional degeneracy here that has nothing to do with the Fermi surface, and that may be reasonably expected to persist even if the Hamiltonian is modified and the Fermi surface is destroyed in one of the ways described above. This is the twofold degeneracy that comes from the fact that two different spin structures have the same (globally minimal) vacuum energy $\E_0(\b s)$.

This twofold degeneracy is topological in origin. For example, it persists to all orders in the strong coupling expansion, since the dispersion in \eqref{def eps(k) lambda} satisfies $\eps_{\b k + \frac12 \b e_1} = \eps_{\b k + \frac12 \b e_2}$. (In fact, at very weak coupling, corresponding to $\lambda \rar \infty$, all four spin structures become degenerate, giving rise to the fourfold ground state degeneracy of the toric code.) Another way to reinforce this point is to note that when $N$ is odd, the number of unpaired fermion modes in each sector changes so that $\Pbb^0(\b s)$ is nonempty for all $\b s$, but nevertheless the $\b s = \frac12 \b e_i$ sectors retain exactly degenerate vacuum energies.

It is thus tempting to connect this with the ground state degeneracy of a U(1)$_2$ CS theory. The latter being a non-spin TQFT can further be connected with the fact that each of these two states corresponds to a different spin structure on the dual side --- in other words, one must ``sum over spin structures'' on $\Mbb$ in order to see this degeneracy.

Nevertheless, it is important to stress that this connection to CS theory is still conjectural. This paper has not demonstrated a deformation of \eqref{def H KS strong} that breaks time-reversal invariance and gaps out the Fermi surface excitations while preserving the topological degeneracy. In fact, a na\"ive attempt to do so --- adding a term of the form $m \sigma^x$ to the matrix $D_{\b k}$ in \eqref{def D} --- does not correspond to a local interaction in the original position space. It is very likely that the needed terms will be versions of Haldane's complex next-nearest-neighbor hopping operators on the honeycomb lattice \cite{Haldane:1988}, or perhaps additional nearest-neighbor hopping operators with position modulation \`a la Kawamoto-Smit. If such a gapped theory is found, the natural next step will be to show that its excitations are semions, as appropriate for a U(1)$_2$ theory.

\section*{Acknowledgments}

It is a pleasure to thank Shaokai Jian, Matt Headrick, and Brian Swingle for useful comments. This work was completed with the support from the Simons Foundation through \emph{It from Qubit: Simons Collaboration on Quantum Fields, Gravity, and Information}, and from the Department of Energy Office of High-Energy Physics grant DE-SC0009987 and QuantISED grant DE-SC0020194. Part of this work was completed during the long-term workshop ``Topological properties of gauge theories and their applications to high-energy and condensed-matter physics'' held at the Galileo Galilei Institute in Arcetri, Italy.

\bibliographystyle{ssg}
\bibliography{Refs}

\end{document}